\documentclass[twocolumn,showpacs,preprintnumbers,amsmath,amssymb]{revtex4}
\usepackage{graphicx}
\usepackage{dcolumn}
\usepackage{epsfig}
\usepackage{bm}
\def\beq{\begin{equation}}
\def\eeq{\end{equation}}
\def\be{\begin{equation}}
\def\ee{\end{equation}}
\def\bea{\begin{eqnarray}}
\def\eea{\end{eqnarray}}

\begin{document}

\title{A precise analytical 
description of the Earth matter effect on oscillations of 
low energy neutrinos} 

\author{A. N. Ioannisian$^{a,b}$, N. A.  Kazarian${^b}$, A. Yu. Smirnov$^{c}$, 
D. Wyler$^d$} 
 \affiliation{
$^a$ Yerevan Physics Institute, Alikhanian Br.\ 2, 375036 Yerevan, 
Armenia\\
$^b$ Institute for Theoretical Physics and Modeling, 375036 Yerevan, 
Armenia\\
$^c$ ICTP, Strada Costiera 11, 34014 Trieste, Italy \\
$^d$  Institut f\"ur Theoretische Physik, Universit\"at Z\"urich,
Winterthurerstrasse 190, CH-8057 Z\"urich, Switzerland}

\begin{abstract}
We present a formalism for the matter 
effects in the Earth on low energy neutrino fluxes which is both 
accurate and has all advantages of a full analytic treatment.
The oscillation probabilities are calculated up to
second order term in $\epsilon(x) \equiv 2V(x)E/\Delta m^2$ where $V(x)$ is
the neutrino potential at position $x$. We show
the absence of large undamped phases which makes the expansion
in $\epsilon$ well behaved.
An improved expansion is presented
in terms of the variation of $V(x)$ around a suitable mean value
which allows to treat energies up to those relevant for Supernova neutrinos.
We discuss also the case of three-neutrino mixing.
\end{abstract}


\maketitle
\section{Introduction}

The propagation of low energy neutrinos in the 
Earth \cite{w,MS86,DN} is an important aspect of physics of 
solar~\cite{w} - \cite{sk-dn} and  
Supernova (SN) neutrinos~\cite{DS} - \cite{DKRT}. 
It will be useful in determining  the 
oscillation parameters, and, in future, to search for effects of 1-3 
mixing~\cite{ohl} and for a 'tomography' of the Earth 
(see, {\it e.g.} \cite{IS,LOTW}). It might even be possible to look
for small structures of the 
density profile~\cite{IS}. 

In the existing calculations of Earth matter effects (see, {\it e.g.} 
~\cite{w} - \cite{IS}) the density profile is often approximated by 
one, two or several layers (mainly mantle and core) 
with  constant densities or a direct numerical integration of 
the evolution equation is performed. However, 
the emergence of the large mixing MSW solution 
to the solar neutrino problem opens a more efficient approach
to the oscillation effects in the Earth. 
Indeed, for the LMA  parameters, 
the oscillations of the  solar and (lower energy) supernova 
neutrinos inside the Earth 
occur in a 'weak' regime,  
where the matter potential $V$ is much smaller than 
the 'kinetic energy' of the neutrino system, {\it i.e.}  
\be
V(x) \ll \frac{\Delta m^2}{2E}. 
\label{cond}
\ee
Here $V(x) \equiv \sqrt{2} G_F N_e(x)$, 
$G_F$ is the Fermi constant,  $N_e(x)$ is the number density of the electrons, 
$\Delta m^2 \equiv m^2_2 - m^2_1$ is the mass squared difference, and $E$ 
is 
the neutrino energy.

In this case one  can introduce a small parameter 
\bea 
\epsilon(x) &\equiv& {2 E V(x) \over \Delta m^2 } 
\label{eps}
\\
&=&0.02 \cdot \left[ { E \over 10 \,  {\rm MeV}}\right]
\cdot \left[  { N_e(x) \over N_A }\right]
\cdot \left[{7.7 \cdot 10^{-5} \, {\rm eV}^2 \over \Delta m^2}\right] ,
\nonumber
\eea
where $N_A$ is the Avogadro number, 
and consider an expansion of the oscillation probabilities in $\epsilon(x)$. 

In ref.~\cite{Ioannisian:2004jk}, 
the $\epsilon$ perturbation theory was formulated
in the basis of neutrino mass states  
$\nu_{mass} \equiv (\nu_1, \nu_2)^T$. 
The oscillation probabilities and the regeneration factor were
calculated to first order in $\epsilon$. The expressions obtained are
valid for  arbitrary density profiles with sufficiently low density  
(\ref{cond}). They simplify the numerical 
calculations substantially and  
allow to understand in details all features of the oscillation effects. 
The method reproduced immediately the analytic result 
obtained in~\cite{HLS} for an approximate but realistic density profile. 
Similar integral expression for the regeneration factor has been discussed  
in~\cite{Akhmedov:2004rq}.

Since $\epsilon(x)$ increases with energy, 
the lowest approximation in $\epsilon(x)$ 
may not be enough for larger energies. For instance, if $E \simeq 50$ MeV 
(possible for SN neutrinos), we find $\epsilon(x) \simeq 0.6$ at 
the center of the Earth.  

The purpose of this paper is to improve on this method and obtain 
accurate formulas which are valid for higher energies.  
In section 2 the oscillation probabilities are calculated in 
second order in $\epsilon(x)$ and the convergence of the $\epsilon$
expansion is commented on. 
In section 3 we suggest an improved perturbation theory 
which allows one to extend the expansion to higher energies. The 
generalization to three neutrinos is given in section 4 and a brief 
conclusion in section 5.

\section{Second order corrections to the oscillation probabilities}

In this and the following section we consider the mixing of
two (active) neutrinos
$\nu_f = U(\theta) \nu_{mass}$, 
where 
$\nu_f \equiv (\nu_e, \nu_a)^T$  and 
$\nu_{mass} \equiv (\nu_1, \nu_2)^T$   
are the flavor and mass states, respectively and
$\nu_a$ is a linear combination of $\nu_{\mu}$ and $\nu_{\tau}$.   
$U(\theta)$ and $\theta$ are the mixing matrix and mixing angle in 
vacuum. We define the matrix $U(\alpha)$ as  
\begin{equation} 
U(\alpha)  \equiv \left( \begin{tabular}{rr} 
$\cos \alpha$ & $\sin \alpha$ \\
$ - \sin \alpha$ & $\cos \alpha$     
\end{tabular}
\right).  
\label{matrix1}
\end{equation}


In ~\cite{Ioannisian:2004jk}  
the following expression for the 
$S$-matrix in the mass eigenstates basis was derived
\footnote{This 
result may be obtained via ordinary 
perturbation theory with the Hamiltonian 
${\cal H}(x)  = diag(0,\Delta) +  
U^\dagger diag(V(x),0) \ U =
{\cal H}^0   +\Upsilon $, where ${\cal H}^0$ is the 
diagonalized Hamiltonian (MSW solution) at point ($x$). 
We would like to stress that only that separation of the Hamiltonian 
into a non-perturbative (${\cal H}^0$)  
and a perturbative ($\Upsilon$) parts leads to results 
where terms proportional to the full distance 
traveled by neutrinos in matter are absent; this is in fact
guaranteed by the existence of the MSW solution.
}:

\bea
S &=&
                        \left( \! \begin{tabular}{cc}
       1 & 0 \\
       0 & $  e^{\! \! \! -i \phi^m_{x_0 \to x_f}}$
                        \end{tabular} \! \! \! \!
                        \right) + \nonumber
\\
&&\hspace{-2cm} - i \int_{x_0}^{x_f}  dx
\left( \! \begin{tabular}{cc}
       1 & 0 \\
       0 & $  e^{\! \! \! -i \phi^m_{x \to x_f}}$
                        \end{tabular}  \! \! \! \!
                        \right)
                        \Upsilon(x)
                        \left( \! \begin{tabular}{cc}
       1 & 0 \\
       0 & $  e^{\! \! \! -i \phi^m_{x_0 \to x}}$
                        \end{tabular} \! \! \! \!
                        \right) -
\nonumber
\\
 && \hspace{-2cm} - \!
                         \int_{x_0}^{x_f} \! \! \! \! \! dx
                         \int_{x_0}^x \! \! \! \! \! dy
                         \left( \! \begin{tabular}{cc}
       1 & 0 \\
       0 & $  e^{\! \! \! -i \phi^m_{x \to x_f}}$
                        \end{tabular}  \! \! \! \!
                        \right)
                        \Upsilon(x)
                        \left( \! \begin{tabular}{cc}
       1 & 0 \\
       0 & $  e^{\! \! \! -i \phi^m_{y \to x}}$
                        \end{tabular} \! \! \! \!
                        \right)
                        \Upsilon(y)
                        \left( \! \begin{tabular}{cc}
       1 & 0 \\
       0 & $  e^{\! \! \! -i \phi^m_{x_0 \to y}}$
                        \end{tabular} \! \! \! \!
                        \right)
\nonumber
\\
&& \hspace{-2cm} +\cdots ,
\label{smat}
\eea
where
\be 
\phi^m_{x_1 \to x_2} \equiv  \int_{x_1}^{x_2} d x \Delta^m (x) 
\label{phim}
\ee
is the adiabatic phase difference acquired by the 
neutrino eigenstates in matter on their trajectory between two 
points $x_1$ and $x_2$.  
$\Delta^m(x)$ is defined as
\be 
\Delta^m(x)  \equiv {\Delta m^2 \over 2E}
           \sqrt{1-2\epsilon(x)\cos 2\theta +\epsilon(x)^2} \ ;
\label{delta}
\ee
in vacuum we obviously have
\be
\Delta^m \rightarrow \Delta \equiv  {\Delta m^2\over 2E}. 
\ee
The $S$-matrix in (\ref{smat}) is written as a perturbative expansion in 
$\Upsilon(x)$  where
\be
\hspace{-0.4cm}
\Upsilon(x) = {\sin 2\theta \over 2} ~V(x)
\left(
\begin{tabular}{cc}
0 & 1 \\
1 & 0
\end{tabular}
\right) + {1 \over 2} \Delta^m (x)\sin^2 \theta' 
\left(
\begin{tabular}{cc}
1 & 0 \\
0 & -1
\end{tabular}
\right).
\label{ups}
\ee
$\theta'$ is the mixing angle of the mass eigenstates in matter, 
\be
\sin 2 \theta^\prime \! = {\epsilon \ \sin 2 \theta \over
\sqrt{(\cos 2 \theta - \epsilon)^2 \! \! + \! \sin^2 2\theta}} =
\epsilon \sin  2\theta^m,
\label{angle}
\ee
and $\theta^m=\theta+\theta^\prime$ 
is the corresponding mixing angle of the flavor states.

The $S$-matrix in eq.(\ref{smat}) refers to a straight path through the earth
from the entry point $x_0$ to an exit point $x_f$ and the coordinate $x$ is measured along the path. For notational 
convenience, we do not put labels $x_0, x_f$, {\it etc.} on $S$.

Using eq. (\ref{ups}), we obtain 
the $S$ matrix in terms of the potential $V$: 
\bea
S \! \! \!  &=& \! \! 
\left( \! \begin{tabular}{cc} 
       1 & 0 \\
       0 & $  e^{\! \! \! -i \phi^m_{x_0 \to x_f}}$ 
                        \end{tabular} \! \! \! \!
                        \right) 
-  
                        i {\sin 2\theta \over 2} 
\! \! \int_{x_0}^{x_f} 
\! \! \! \! \! \! \! dx  V(x) \! \!  
                        \left( \! \! \begin{tabular}{cc} 
                    \hspace{-0.5cm}        0 & 
\hspace{-0.5cm} $e^{\! \! \! -i \phi^m_{x_0 \to x}}$ \\
                            $e^{\! \! \! -i \phi^m_{x \to x_f}}$ & \hspace{-0.5cm}   0  
                        \end{tabular} \! \! \! \! \right)
\nonumber
\\
&& \hspace{-0.7cm} -
                        i {\sin^2 2 \theta \over 4 \Delta} 
                                \left( \!  \begin{tabular}{cc}
       $ 1 $ & $0$ \\
       $0$ & $  -e^{\! \! \! -i \phi^m_{x_0 \to x_f}}$
                        \end{tabular} \! \! \! \!
                        \right)  \! \!
\int_{x_0}^{x_f} \! \! \! \! \! \! \! dx \cdot V(x)^2
\label{A}
\\
&& \hspace{-0.7cm} -
                        {\sin^2 2 \theta \over 4} \int_{x_0}^{x_f}
			\! \! \! \! \! \! \! dx
                        \int_{x_0}^x 
			\! \! \! \! \! dy \ V(x) V(y)
                        \left( \! \begin{tabular}{lc}
      $ e^{\! \! \! -i \phi^m_{y \to x} } $ & $ \hspace{-0.4cm} 0$ \\
      $ 0$ & $  \hspace{-0.4cm}e^{\! \! \! -i \phi^m_{x_0 \to x_f} \! \! \! 
                   +i \phi^m_{y \to x} }$
                        \end{tabular}\! \! \! \!
                        \right) .
\nonumber
\eea
The two last terms (proportional to $\epsilon^2$) come from 
the first order
in $\Upsilon$ (term proportional to  $\sin^2 \theta'$ in Eq. (\ref{ups}))  
and the second order in $\Upsilon$ (see Eq. (\ref{smat})) correspondingly.

Using the evolution matrix in the mass state basis (\ref{A}), we can 
calculate the amplitudes  and probabilities of various transitions. 
The evolution matrix from the mass states to  the flavor states  
relevant for the solar and SN neutrinos equals
$U S$, 
where $U$ is the vacuum mixing matrix (\ref{matrix1}). 
Consequently, the amplitude of the mass-to-flavor transition, 
is given by 
\be
A_{\nu_i \to \nu_\alpha} = U_{\alpha j }(\theta) S_{ji}. 
\ee

The  probability of the $\nu_2 \to \nu_e$ transition,  
$P_{\nu_2 \to \nu_e} = |A_{\nu_2 \to \nu_e}|^2 = 
|U_{e j }(\theta) S_{j2}|^2$ is then found to be    
\bea
\label{p2}
\hspace{-0.3cm}
P_{\nu_2 \to \nu_e} &=&        
      \ \sin^2 \theta + {1 \over 2} \sin^2 2 \theta 
      \int_{x_0}^{x_f} \! \! \! \! \! dx  
      \, V(x) \sin \phi^m_{x \to x_f}
\nonumber
\\
&& \hspace{-1.5cm} + {1 \over 4} \, \sin^2 2 \theta \, \cos 2 \theta
        \int_{x_0}^{x_f} \! \! \! \! \! dx \int_{x_0}^{x_f}\! \! \! \! \! dy
         V(x)  \, V(y)  \, \cos \phi^m_{y\to x}, 
\label{pord}
\eea
where the last term is  the $\epsilon^2$ correction. 
The integrations over $x$ and $y$ can be disentangled. Indeed,  
writing $\phi^m_{y\to x} = \phi^m_{y\to z} + \phi^m_{z \to x}$, 
where $z$ is an arbitrary point of the trajectory,  we find 
\bea
\label{split}
&& \hspace{-1cm} 
 \int_{x_0}^{x_f} \! \! \! \! \! dx \int_{x_0}^{x_f}\! \! \! \! \! dy
         V(x)  \, V(y)  \, \cos \phi^m_{y\to x}
=
\\
&&\left[ \int_{x_0}^{x_f} \! \! \! \! \! dx \, V(x) \cos \phi^m_{z \to x}
\right]^2
 +
\left[ \int_{x_0}^{x_f} \! \! \! \! \! dx \, V(x) \sin \phi^m_{z \to x}
\right]^2 .
\nonumber
\eea
This shows that the second order correction is positive for all $V$ which
do not vanish. 

Furthermore, for  a symmetric density 
profile (with respect to the middle point of the 
trajectory) the second term in 
(\ref{split}) vanishes. 
This can be seen immediately by   
choosing $z=\bar{x}\equiv (x_f+x_0)/2$ in the center of the trajectory. 
So, finally we obtain  for a symmetric profile
\bea
P_{\nu_2 \to \nu_e} &=& \sin^2 \theta
\nonumber
\\
&+& {1 \over 2} \, \sin^2 2 \theta \,
    \int_{x_0}^{x_f} \!  \!  dx \, V(x)  \, \sin \phi^m_{x \to x_f}
\nonumber
\\
&+& {1 \over 4} \sin^2 2 \theta \, \cos 2 \theta
        \left[ \int_{x_0}^{x_f} \! \! \! \! \! dx \,
         V(x)  \cos \phi^m_{\bar{x} \to x} \right]^2
\eea
or (using again the symmetry of $V$)
\bea
P_{\nu_2 \to \nu_e}&=& \sin^2 \theta
\nonumber
\\
&+& {1 \over 2} \, \sin^2 2 \theta \, \sin \phi^m_{\bar{x} \to x_f}
    \int_{x_0}^{x_f} \! \! \! dx \, V(x)  \, \cos \phi^m_{\bar{x} \to x}
\nonumber
\\
&+& {1 \over 4} \sin^2 2 \theta \, \cos 2 \theta
        \left[ \int_{x_0}^{x_f} \! \! \! \! \! dx \,
         V(x)  \cos \phi^m_{\bar{x} \to x} \right]^2.
\label{pii}
\eea
The phase $\phi^m_{\bar{x} \to x_f}$ should be calculated according to 
(\ref{phim}).  

The two last terms in (\ref{pii}) determine  the regeneration 
parameter defined as $f_{reg} \equiv P_{\nu_2 \to \nu_e} -\sin^2 \theta$ 
(see, {\it e.g.}, \cite{GCS}).  
The probability of the  $\nu_1 \to \nu_e$ oscillations 
can be obtained immediately from the unitarity condition 
$P_{\nu_1 \to \nu_e}  =   1 - P_{\nu_2 \to \nu_e}$.

According to (\ref{pii}) the effective expansion parameter of the series  
is 
\vspace{-0.1cm}
\be
I \equiv \int_{\bar{x}}^{x_f} 
\! \! \! dx \, V(x)  \cos \phi^m_{\bar{x} \to x} , 
\label{I}
\ee
so that 
\be
\hspace{-0.4cm}
P_{\! \nu_2 \to \nu_e} \! \! \! =  \sin^2 \theta + \sin^2 2 \theta 
\left[ \sin \phi^m_{\bar{x} \to x_f} I +  \cos 2\theta  I^2  + ...\right]. 
\label{series}
\ee
Notice that here the adiabatic phase should be calculated from the center 
of trajectory to a given point $x$, which corresponds to the explicit 
analytic expression obtained in Ref.~\cite{HLS}.  
According to (\ref{series}) the first order correction is absent for 
trajectories with $\phi^m_{\bar{x} \to x_f} = \pi k$, ($k$ = integer) and 
the second order correction would be zero for maximal vacuum mixing.  

Taking $\Delta_m \approx \Delta $ we obtain the useful bound 
\be
I \sim \frac{2E}{\Delta m^2} 
\int_{y(\bar{x})}^{y(x_f)} \! \! \! \! dy  V(y) \cos y \leq  
\frac{2EV_{max}}{\Delta m^2} = \epsilon_{max}.
\ee
$V_{max}$ is the maximum value of the potential on the trajectory
and $y(x) = \frac{\Delta m^2}{2E}x$.

In  eq.(\ref{A}) we note the presence of a possibly large phase 
$\phi^m_{x_0 \to x_f}$ and an undamped integral
in the term $\sim V(x)^2$ (see 1-1 element of the matrix).  It  originates 
from  $\epsilon^2$ term in $\Upsilon$. 
(The undamped terms are absent in the linear term in $\epsilon$ 
\footnote{It is important to recall that $\epsilon$ enters 
both through $V$ and through
the adiabatic phase $\phi^m$.}.)
This could be a problem, because the potential (squared)  is integrated 
over a large distance without an oscillatory damping, and this might give rise
to a large second order term in the expansion. However by a simple partial 
integration 
of the last,  $\sim V(x)V(y)$,  term in (\ref{A}) one can see that the 
undamped integral cancels.  We have verified that this also
happens in order  $V^3$ for constant potentials. Therefore
the  $\epsilon$ expansion appears to be well behaved (see also [26]).

\section{Improved perturbation theory}

As mentioned before, the accuracy of our expressions decreases 
for higher 
densities and energies.  However, the expansion parameter can be reduced
and therefore the expansion can be improved. 
This can be achieved by
considering a perturbation around some average potential $V_0$ rather 
than around the vacuum value $V_0 = 0$ \footnote{Even more general would 
be
an expansion around a suitable potential for which there is a closed
analytic form.}. In this case we expect 
the expansion parameter to be 
\be 
\epsilon = \frac{2E\Delta V}{\Delta m^2} = \frac{2E(V - V_0)}{\Delta m^2}. 
\ee
The corresponding results can be immediately obtained from the
original perturbation theory. Indeed, the transition to an
average potential $V_0$ is equivalent to 
considering the  problem in the basis $\nu^0_m =
(\nu_{1}^0, \nu_{2}^0)$,  
where $\nu_{i}^0$ are the eigenstates of the Hamiltonian in matter with 
a constant potential $V_0$. These states are analogous to mass 
eigenstates in the $V_0=0$ theory. Therefore the $S$-matrix $S^0$ for  
$(\nu_{1}^0, \nu_{2}^0)$ follows from the  $S$ matrix for mass eigenstates 
(\ref{A}) by the substitution 
\be
V \rightarrow  \Delta V \equiv V - V_0, ~~~ \theta \rightarrow 
\theta_0^m , 
\ee
where 
$\theta_0^m$ is the flavor mixing angle in matter with 
the potential $V_0$:  
\be 
\sin 2 \theta^m_0 = 
{\sin 2 \theta \over \sqrt{1-2 \epsilon_0 \cos 2 \theta + \epsilon_0^2} }
\ee
and 
\vspace{-0.2cm}
\be
\epsilon_0 \equiv  {2 E V_0 \over \Delta m^2}. 
\ee
The adiabatic phase differences 
generated for the eigenstates 
traveling in matter with true $V$ 
are invariant under a shift of the average potential,  so that 
the phases, $\phi^m_{x_i \to x_j}$ 
are unchanged. Therefore 
\vspace{-0.2cm}
\be
S^0 = S(\Delta V, \theta_0^m). 
\label{s0}
\ee

We introduce the mixing matrix 
\be
U_0^\prime \equiv U(\theta_0^\prime)
\ee 
which relates 
the eigenstates of neutrinos in the potential 
V$_0$ to the mass eigenstates in vacuum: $\nu_{mass} = U_0^\prime 
\nu_m^0$. 
The angle $\theta_0^\prime$ 
is given by 
\be 
\sin 2 \theta^\prime_0 = \epsilon_0 \sin 2 \theta^m_0
\ee
and it is easy to check that 
$\theta= \theta^m_0 - \theta^\prime_0$. 

Now the amplitude of the mass-to-flavor transition, $\nu_i \to \nu_\alpha$,
equals 
\vspace{-0.1cm}
\be
A_{\nu_i \to \nu_\alpha} = U_{\alpha j }(\theta^m_0) (S^0)_{jk} 
U^{\dagger}_{ki} 
(\theta^\prime_0). 
\ee

A straightforward calculation leads to 
the $\nu_2 \to \nu_e$ oscillation probability 
$P_{\nu_2 \to \nu_e} = |A_{\nu_2 \to \nu_e}|^2$

\begin{widetext}
\bea
\hspace{-0.2cm}P_{\nu_2 \to \nu_e} &=& 
\sin^2 \theta +\epsilon_0 \, \sin^2 2 \theta^m_0
\, \sin^2 {\phi^m_{x_0 \to x_f} \over 2}
+ {1 \over 2} \, \sin^2 2 \theta^m_0 \, \cos 2 \theta^\prime_0
   \int_{x_0}^{x_f} \! \! \! dx \, \Delta V(x) \, \sin \phi^m_{x \to x_f}
\nonumber
\\
&+& {\epsilon_0 \over 2} \, \sin^2 2 \theta^m_0 \, \cos 2 \theta^m_0
   \int_{x_0}^{x_f} \! \! \! dx \, \Delta V(x) \, \sin \phi^m_{x_0 \to x}
-{\epsilon_0 \over 8} \sin^4 2 \theta^m_0 \! \! 
     \int_{x_0}^{x_f} \! \! \! \! \! dx \int_{x_0}^{x_f}\! \! \! \! \! dy
        \Delta V(x)  \, \Delta V(y) 
            \cos (\phi^m_{x_0 \to x} \! \! \! - \! \phi^m_{y \to x_f})
\nonumber
\\
&+& {1 \over 8} \, \sin^2 2 \theta^m_0 
        (\cos 2 \theta^m_0 + \cos 2 \theta^\prime_0 -2 \sin^2 \theta 
                -\epsilon_0 \sin^2 2 \theta ^m_0 ) 
\int_{x_0}^{x_f} \! \! \! \! \! dx \int_{x_0}^{x_f}\! \! \! \! \! dy
         \Delta V(x) \, \Delta V(y) \cos \phi^m_{y\to x} \ .
\label{pimp}
\eea
\end{widetext}

We note that there are two first order (in $\Delta V$) terms, one 
containing  
$\phi^m_{x_0 \to x}$, the other $\phi^m_{x \to x_f}$ in contrast to the 
original theory which contains the phase $\phi^m_{x \to x_f}$ only.   

For $V_0=0$ eq. (\ref{pimp}) coincides with the 
previous result (\ref{pord}). 

For  a symmetric density profile we obtain

\begin{widetext}
\bea
\hspace{-0.3cm} 
P_{\nu_2 \to \nu_e}
\! \! &=& \! \! \sin^2 \theta +\epsilon_0 \, \sin^2 2 \theta^m_0
\, \sin^2 \phi^m_{\bar{x} \to x_f} 
+ {1 \over2}\sin^2 2 \theta^m_0 \, 
           ( \cos 2 \theta^\prime_0 +\epsilon_0 \cos 2\theta^m_0 )
       \sin \phi^m_{\bar{x} \to x_f}
\int_{x_0}^{x_f} \! \! \! dx \, \Delta V(x) \, 
                \cos \phi^m_{\bar{x} \to x} 
+
\nonumber
\\
&& 
+{1 \over 8} \, \sin^2 2 \theta^m_0 
        (\cos 2 \theta^m_0 + \cos 2 \theta^\prime_0 -2 \sin^2 \theta 
                -2 \epsilon_0 \sin^2 2 \theta ^m_0 )  
\left[ \int_{x_0}^{x_f} \! \! \! \! \! dx 
         \Delta V(x) \,\cos \phi^m_{\bar{x} \to x}\right]^2. 
\label{symp} 
\eea
\end{widetext}
\noindent

Thus the effective expansion parameter of the series in the 
improved perturbation theory is 
\be
\int_{\bar{x}}^{x_f} \! \! \! dx \, \Delta V(x) \, 
                \cos \phi^m_{\bar{x} \to x} .
\label{iI}
\ee

The choice of V$_0$ is arbitrary; 
the full expansion of the $S$ matrix does not depend on it.
It just should be chosen in a clever way. 

To illustrate the improvements, let us consider  neutrinos with 
energy 50 MeV [100 MeV]. 
For such neutrinos $\epsilon$=0.2 [0.4] in the 
upper mantle and $\epsilon$=0.6 [1.2] in the core. Thus,  the average is 
${\epsilon_0} \simeq 0.4$ [0.8]. Without improvement, one
expects the accuracy of the computation of $\epsilon^3 \simeq 0.2$ [$O(1)$] 
in the core; with the improvement it is  reduced to 
$(\epsilon -\epsilon_0)^3 \simeq 0.01$ [0.06].   
The optimal $V_0$ can be chosen independently for each trajectory 
inside the Earth. For a mantle crossing trajectory, for instance, one
would take the average value in the mantle. 

\begin{figure}[t]
\begin{center}
\epsfig{file=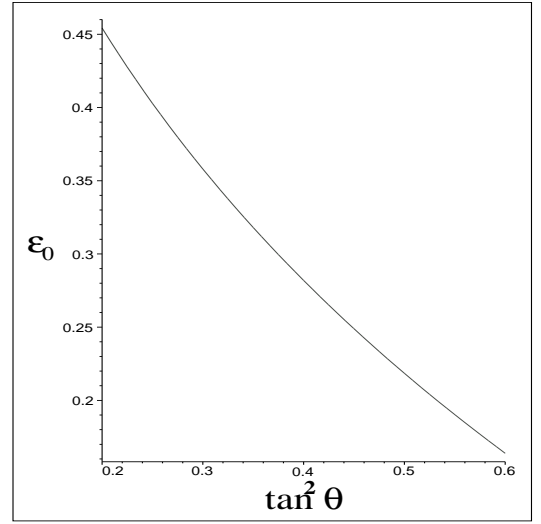,width=0.8 \linewidth,height=.8 \linewidth}
\caption{\label{fig1} The dependence of $\epsilon_0 =2EV_0/ \Delta m^2 $ 
on $\tan^2 \theta$ obtained by setting the prefactor in eq. (\ref{prefactor}) 
equal to zero.}
\end{center}
\vspace{-1cm}
\end{figure}

A 'good' value of $\epsilon_0$ may come from the observation that
the second order term in (\ref{symp}) is multiplied by the  prefactor 
\be
(\cos 2 \theta^m_0 + \cos 2 \theta^\prime_0 -2 \sin^2 \theta 
                -2 \epsilon_0 \sin^2 2 \theta ^m_0 ) \ .
\label{prefactor}
\ee
Since $\epsilon_0$ is arbitrary, one may choose it such 
that this prefactor vanishes. 
In Fig.1  we show  the $\tan^2 \theta$ dependence of 
$\epsilon_0$ for which the prefactor vanishes.

In the limit $V \rightarrow 0$ the second, the third and the forth terms 
in (\ref{symp}) cancel  each other ( up to $\epsilon_0^3$),  
and the probability reduces to $\sin^2 \theta$.

\section{Corrections due to  three-neutrino mixing}

In the standard parametrization the lepton mixing 
matrix is 
\bea
&& \hspace{-0.5cm} U = O_{23} \, diag(1,1,e^{i\delta_{cp}}) O_{13} 
\, diag(1,1,e^{-i\delta_{cp}}) O_{12} = 
\nonumber
\\
&& \hspace{-0.8cm} 
{ 
\left( \! \! \! \!
\begin{tabular}{ccc}
$c_{13} c_{12}$ & $c_{13} s_{23}$ & $s_{13} e^{-i\delta_{cp}}$ \\
$-s_{12} c_{23} - c_{12} s_{23} s_{13} e^{i\delta_{cp}}$ & 
   $c_{12} c_{23}-s_{12}s_{23}s_{13}e^{i\delta_{cp}}$ & $c_{13}s_{23}$\\
$s_{12}s_{23}-c_{12}c_{23} s_{13} e^{i\delta_{cp}}$ &  
   $-c_{12} s_{23} - s_{12}c_{23} s_{13} e^{i\delta_{cp}}$ &
         $c_{13}c_{23}$ 
\end{tabular} \! \! \!
\right)
}
\label{mmu}
\nonumber
\eea

By a redefinition of the mixing matrix
\be
U \to U \cdot diag(1,1,e^{i\delta_{cp}})
\label{unew}
\ee
the Hamiltonian becomes real, {\it i.e.}
\bea
\hspace{-1cm}{\cal H} &=&  
\left(
\begin{tabular}{ccc}
0 & 0 & 0 \\
0 & $\Delta_s$ & 0 \\
0 & 0 & $\Delta_a$
\end{tabular}
\right) + U^\dagger 
\left(
\begin{tabular}{ccc}
V & 0 & 0 \\
0 & 0 & 0 \\
0 & 0 & 0
\end{tabular}
\right)
U 
\\
&=&
\left(
\begin{tabular}{lll}
$V c^2_{13} c^2_{12}$ & 
    \ $ V c^2_{13} s_{12} c_{12}$ & 
    \    $ V c_{12} c_{13} s_{13}$ \\
$V c^2_{13} s_{12} c_{12} $ & 
    \ $\Delta_s +V c^2_{13} s^2_{12}$ & 
    \    $ V s_{12} c_{13} s_{13}$ \\
$V c_{12} c_{13} s_{13}$ & 
    \ $V s_{12} c_{13} s_{13}$ & 
    \    $\Delta_a + V s^2_{13}$
\end{tabular}
\right),
\label{3nh}
\eea
where $\Delta_s \equiv \Delta m^2_{\odot} / 2E$ and 
$\Delta_a \equiv \Delta m^2_{atm} / 2E -\Delta_s$.

Thus we see that both the CP phase $\delta_{cp}$ and $\theta_{23}$
do not influence the propagation in matter (determined by the Hamiltonian).
Also, since in (\ref{unew}) the first line does not contain 
$\delta_{cp}$  and $\theta_{23}$ these parameters disappear
in the oscillations from $\nu_e$ to $\nu_e$, or from 
$\nu_e$ to mass eigenstates and vice versa.
They manifest themselves only when one considers the flavor
states $\nu_\mu$ or $\nu_\tau$.

These arguments are general and are valid in arbitrary matter density.
 
We now write the Hamiltonian in the form
\be
{\cal H}={\cal H}_{(3\nu)}^0 +   \Upsilon_{(3\nu)} ,
\ee
where
\be
{\cal H}_{(3\nu)}^0 = 
\left(
\begin{tabular}{ccc}
0 & 0 & 0 \\
0 & $\Delta_s^m$ & 0 \\
0 & 0 & $\Delta_a^m$
\end{tabular} 
\right)
\ee
and
\bea \hspace{-0.4cm}
\Upsilon_{(3\nu)} \! \! \! &=& \! \!    
{\cal H} \! - \! {\cal H}^0_{3\nu} \! + \! diag(0,\Delta_s , \Delta_a) -
{V \! \!  +  
\! \! \Delta_s \! \! +  
\! \!\Delta_a  \!  \!- \! \! \Delta_s^m  
\! \! \! \! - \! \! \Delta_a^m \over 3} I
\nonumber
\\
&&\hspace{-1.4cm}=
V \, c_{13}^2
\left(
\begin{tabular}{ccc}
  0               & $\sin 2 \theta_{12} /2 $ & $c_{12} s_{13}/c_{13} $ \\
$\sin 2 \theta_{12} /2$ & 0                   & $s_{12} s_{13}/c_{13} $ \\
$c_{12} s_{13}/c_{13}$      & $s_{12} s_{13}/c_{13}$ & 0
\end{tabular}
\right) \! + \! O(V^2).
\label{subst}
\eea
$\Delta_s^m$  and $\Delta_a^m$  are the 
eigenvalues of the Hamiltonian in matter~\footnote{ 
When $V \ll \Delta_s \ll \Delta_a$ then  
$\Delta^m_a \simeq \Delta_a + O(V)$ and  \\
$\Delta^m_s \simeq \Delta_s 
\sqrt{(\cos 2 \theta -{V \, c^2_{13}\over \Delta_s} )^2+\sin^2 2 \theta} 
+O(s_{13}^2{V^2 \over \Delta_a})$.  
}.
In eq. (\ref{subst}) we have subtracted a term proportional to the unit
matrix in order to make it
traceless and thus convenient for a power expansion.

A straightforward calculation leads to the transition probabilities of
the mass eigenstates to $\nu_e$: 
\bea
\hspace{-0.3cm}
P_{\nu_1 \to \nu_e} \! \! \! \! \!  &=& \! \! \! 
c_{13}^2 c_{12}^2
-{\sin^2 2 \theta_{12} \over 2} c_{13}^4 
                       \int_{x_0}^{x_f} \! \! \! \! \!  dx \, V  
                                 \sin \phi_{x \to x_f} 
\nonumber
\\
&&\hspace{-0.3cm}- 2 \, c_{12}^2 c_{13}^2 s_{13}^2
\int_{x_0}^{x_f} \! \! \! \! \!  dx \, V  
                                 \sin \psi_{x \to x_f} ,
\\
\hspace{-0.3cm}
P_{\nu_2 \to \nu_e} \! \! \! \! \!  &=& \! \! \! 
c_{13}^2 s_{12}^2
+{\sin^2 2 \theta_{12} \over 2} c_{13}^4 
                       \int_{x_0}^{x_f} \! \! \! \! \!  dx \, V  
                                 \sin \phi_{x \to x_f} 
\nonumber
\\
&& \hspace{-0.3cm}- 2 \, s_{12}^2 c_{13}^2 s_{13}^2
\int_{x_0}^{x_f} \! \! \! \! \!  dx \, V  
                                 \sin (\psi_{x \to x_f}\! \! \! - \!
\phi_{x\to x_f} )  ,
\\
\hspace{-0.3cm}
P_{\nu_3 \to \nu_e} \! \! \! \! \!  &=& \! \! \! 
s_{13}^2 
+ 2 \, c_{12}^2 c_{13}^2 s_{13}^2
\int_{x_0}^{x_f} \! \! \! \! \!  dx \, V  
                                 \sin \psi_{x \to x_f} 
\nonumber
\\
&&\hspace{-0.3cm}+
  2 \, s_{12}^2 c_{13}^2 s_{13}^2
\int_{x_0}^{x_f} \! \! \! \! \!  dx \, V  
                                 \sin (\psi_{x \to x_f}\! \! \! - \!
\phi_{x\to x_f} ) ,  
\eea
where
\be
\phi_{a\to b} = \int_a^b \Delta_s^m(x) \, dx \ ,
~~~~~
\psi_{a\to b} = \int_a^b \Delta_a^m(x) \, dx \ .
\ee
The function $\sin \psi_{x \to x_f}$ oscillates $\Delta_a^m / \Delta_s^m \simeq 
\Delta m^2_{atm}/\Delta m^2_{\odot}$ times faster than   
$\sin \phi_{x \to x_f}$. Thus,  the corresponding integral is roughly 
$\Delta m^2_{atm}/\Delta m^2_{\odot}$ times smaller than the one which contains
the phase 
$\phi$; furthermore,  it has  a prefactor 
$s_{13}^2$. Therefore we get to a good approximation
\bea
P_{\nu_1 \to \nu_e} \! \! \! \! \!  &=& \! \! \! 
c_{13}^2 c_{12}^2
-{\sin^2 2 \theta_{12} \over 2} c_{13}^4 
                       \int_{x_0}^{x_f} \! \! \! \! \!  dx \, V  
                                 \sin \phi_{x \to x_f}  ,
\label{p1e} 
\\
P_{\nu_2 \to \nu_e} \! \! \! \! \!  &=& \! \! \! 
c_{13}^2 s_{12}^2
+{\sin^2 2 \theta_{12} \over 2} c_{13}^4 
                       \int_{x_0}^{x_f} \! \! \! \! \!  dx \, V  
                                 \sin \phi_{x \to x_f}   ,
\\
P_{\nu_3 \to \nu_e} \! \! \! \! \!  &\simeq& \! \! \! 
s_{13}^2 .
\label{p3e}
\eea

These results may be also obtained from 
eq. (\ref{3nh}) \cite{Akhmedov:2004rq} (see \cite{ohl} for some earlier 
discussion).
If $\Delta_a \gg \Delta_s \gg V$ and $s_{13} \ll 1$, the third neutrino 
decouples and one arrives at the two neutrino propagation problem in  
matter with potential $V \to V c_{13}^2$ and mixing angle $\theta_{12}$.
Following the procedure of section II and using the full mixing matrix 
$U \, diag(1,1,e^{i\delta_{cp}})$ we easily recover eqs. 
(\ref{p1e}) -(\ref{p3e}).

\section{Conclusion}

Motivated by the large mixing MSW solution to the solar neutrino we
have developed a simple formulation of the earth matter effects
on low energy neutrino beams. Following
\cite{Ioannisian:2004jk}, we derive an expansion for the
neutrino transitions in terms of the parameter
\vspace{-0.3cm}
$$
\epsilon(x) \equiv {2 E V(x) \over \Delta m^2 } 
$$
to second order. By choosing a convenient constant average value for the 
neutrino potential as starting point, the precision can be substantially
improved and it is possible to reach an accuracy of a few percents even for
energies near $70 - 80$ MeV. The effective expansion parameter is a simple
integral in eq. (\ref{I}) (or eq. (\ref{iI})) 
together with eq. (\ref{phim})) which can be done numerically. 
The expansion allows for
a convenient quantitative discussion of various physical effects such 
as the attenuation effect to the remote structures of the density profile 
or the effect of energy resolution of  detectors. 
We also consider the case of three-neutrino mixing.


The work of A.N.I. was supported by the Swiss National 
Science Foundation (SNF). A.N.I. thanks the University of Z\"urich for 
hospitality.


\end{document}